# A robust DOA estimation method for a linear microphone array under reverberant and noisy environments

*Hao Wang and Jing Lu*

Author Key Laboratory of Modern Acoustics, Institute of Advanced Artificial Intelligence in Nanjing, Nanjing 210014, China

haowang@smail.nju.edu.cn, lujing@nju.edu.cn

## Abstract

A robust method for linear array is proposed to address the difficulty of direction-of-arrival (DOA) estimation in reverberant and noisy environments. A direct-path dominance test based on the onset detection is utilized to extract time-frequency bins containing the direct propagation of the speech. The influence of the transient noise, which severely contaminates the onset test, is mitigated by a proper transient noise determination scheme. Then for voice features, a two-stage procedure is designed based on the extracted bins and an effective dereverberation method, with robust but possibly biased estimation from middle frequency bins followed by further refinement in higher frequency bins. The proposed method effectively alleviates the estimation bias caused by the linear arrangement of microphones, and has stable performance under noisy and reverberant environments. Experimental evaluation using a 4-element microphone array demonstrates the efficacy of the proposed method.

## 1. Introduction

Direction-of-arrival (DOA) estimation using a microphone array plays an important role in speech capturing and processing for human-robot voice interaction, video conferencing, camera steering, and intelligent monitoring. Besides, the location parameters of the speech source are an important priori information for further processing such as speech enhancement, sound source separation, and sound scene analysis. The commonly utilized DOA estimation methods are based on time-delay-of-arrival (TDOA) [1-4], such as the most widely used generalized cross-correlation (GCC) approach [3] to time delay estimation (TDE) [6], steered response power (SRP) [4-8], which forms a conventional beam, scans it over the appropriate region of the working space, plots the magnitude squared of the output, and spatial spectrum [8-12], including the popular multiple signal classification (MUSIC) algorithm [11,12]. In addition to the above three, methods based on deep learning strategy [13,14] have also attracted the researchers' attention recently. Although most of these algorithms perform reliably in free-field conditions, the reverberation and noise in practical application environments often deteriorate the performance of DOA estimations.

Many algorithms have been proposed to improve the robustness of the system in the presence of reverberation and noise. The phase transform (PHAT) weighting was proposed to improve the performance of the popular SRP method, resulting in the well-known SPR-PHAT algorithm [8, 15]. The robust relative transfer function (RTF) estimators [16] were proposed to exploit the speech features under noisy conditions. Recently, a direct-path dominance (DPD) test [17-19] was proposed based on the ratio of the first two eigenvalues of the spatial spectrum matrix, resulting in considerable improvement of estimation accuracy in adverse circumstances with a 32-microphone spherical array. Meanwhile, another DPD test was proposed with an envelope tracking strategy for speech, reverberation and background noise in [20]. Inspired by the precedence effect [21-23], the onset feature [24] can also be classified as a DPD test method.

The attention of this paper is focused on the DOA estimation in noisy and reverberating rooms. To deal with this problem and guarantee stable performance under noisy and reverberant environments, a robust estimation method is proposed as follows. Firstly, a pragmatic DPD test based on the onset feature detection is utilized. A method of determining and eliminating the influence of the transient noise is also launched to further improve the robustness of the DPD test in adverse environments. Secondly, a robust but possibly biased TDOA is estimated in middle frequency range, using a modified weighted prediction error (WPE) method [25-30] to attenuate the influence of reverberation in each extracted bin because human voice has the highest power in this range. Finally, a more precise DOA is obtained by further refinement in higher frequency bins due to this rage contains accurate DOA information.

## 2. Algorithm description

### 2.1. Robust DPD test based on onset feature

The signals received in $I$ channels are transformed into the frequency domain via short time Fourier transform (STFT), and described as $x_i(n,k)$, where $i$ ($i = 1, 2,…, I$) is the channel index, $n$ is the time frame index, and $k$ is the frequency bin. The average amplitude of the cross-power spectrum is calculated as

$$C(n,k) = \frac{1}{I^2} \sum_{i_1=1}^{I} \sum_{i_2=1}^{I} \left| x_{i_1}(n,k) x_{i_2}^*(n,k) \right| \qquad (1)$$

and logarithmically compressed to

$$P(n,k) = \log_{10}(C(n,k) + \xi), \qquad (2)$$

where the regularization term $\xi$ is utilized to reduce the influence of background noise on the stability of the algorithm, the superscript "*" and "$|\cdot|$" denote the complex conjugate operation and the absolute value respectively [20]. $P(n,k)$ can be regarded as the power envelope of the obtained signal in the STFT domain.

Inspired by the precedence effect [23], the onset edge STFT bins of speech can be considered as the direct propagation of the speech signal, which contains the most accurate DOA information. The power envelope suddenly increases at the onset of speeches. To measure the increment, a rate of change of the logarithm of average cross-power spectrum suitable for this algorithm is defined as

$$\Delta P(n,k) = P(n,k) - \frac{1}{N_t}\sum_{t=1}^{N_t} P(n-t,k), \quad (3)$$

where $P(n-t,k)$ is the power envelope of the $t$-th frame before frame $n$ in frequency $k$ and $N_t$ is the number of frames involved in calculating the rate of change. The $K$ bins with the highest rate of change pass the onset DPD test, forming a set identified as

$$\Pi \triangleq \{(n,k) \mid \text{the largest } K \ \Delta P(n,k)\}. \quad (4)$$

It is easy to see that a shorter frameshift can increase the number of selected bins so as to improve the accuracy of DOA estimation.

In actual scenarios, there is always some ambient interference. Common interference noise can be divided into the following categories: steady-state noises, like fan noise and electrical noise; transient noises, like door slam, knocking noise, and keyboard sound; other competing speaker's speech, like music interference and TV-noise. The steady-state noises can be ignored because their sound power does not change quickly. The sound power of target speeches is usually greater than ambient interference, thus compared with the competitor's speech, the target speeches are the main components at the time-frequency bins where the power increases faster. But transient noises have the greatest impact on the DPD test because they have high power rates of change in time-frequency bins and induce considerable misjudgment of the onset test.

This part describes the determination and elimination of transient noise. Transient noises have the characteristics of large power and short time interval, based on which two determination criteria can be designed as follows.

(1) Calculate the total power envelope of each frame as $P_t(n) = \sum_k P(n,k)$, and search the index of the frame with local maximum as

$$n_v = \{n \mid P_t(n+1) - P_t(n) < 0, \ P_t(n) - P_t(n-1) > 0\}. \quad (5)$$

(2) A reasonable measure of the rate of power change can be calculated as

$$\max_{dn}\left(\frac{(P_t(n_v) - P_t(n_v - dn))}{dn}\right) > V_1, \ dn \in \left[1, \frac{\Delta n}{2}\right], \quad (6)$$

$$\max_{dn}\left(\frac{(P_t(n_v) - P_t(n_v + dn))}{dn}\right) > V_2, \ dn \in \left[1, \frac{\Delta n}{2}\right], \quad (7)$$

where $\Delta n$ denotes the range of the local interval, $V_1$ and $V_2$ are the thresholds of the uphill and downhill slope of the power spectrum respectively.

If both of the above two criteria are satisfied, the frame with index $n_v$ is determined to be a transient noise frame, and $\Delta n$ frames centered at $n_v$ will be ignored in the onset DPD test in Eq. 4, rewritten as

$$\Pi \triangleq \{(n,k) \mid \text{the largest } K \ \Delta P(n,k) \cdot \varepsilon(n)\}, \quad (8)$$

where

$$\varepsilon(n) = \begin{cases} 0, & |n - n_v| \leq \dfrac{\Delta n}{2} \\ 1, & |n - n_v| > \dfrac{\Delta n}{2} \end{cases}. \quad (9)$$

Some illustrating examples are presented in Figs. 1 and 2, in which the speech and interference are recorded in three rooms (Room 1: 4.5 × 7.4 × 3 m$^3$, T60 = 0.32 s; Room 2: 3.5 × 5.2 × 3 m$^3$, T60 = 1.2 s; Room 3: 7.35 × 5.9 × 5.22 m$^3$, T60 ≈ 5 s) by a 4-element uniform linear array (ULA) with adjacent interval of 0.035 m. The speaker and the interference are 2 m away from the center of ULA with incident angles of 45° and −45° respectively. The data is sampled at 16 kHz and analyzed with the frame size of 512 points and the frameshift of 8 points. Figure 1 shows the mean amplitude of the cross-power spectrum, the bins that pass the onset DPD test (red "×"s) and the transient noise bins that fail the onset DPD test (black "+"s).

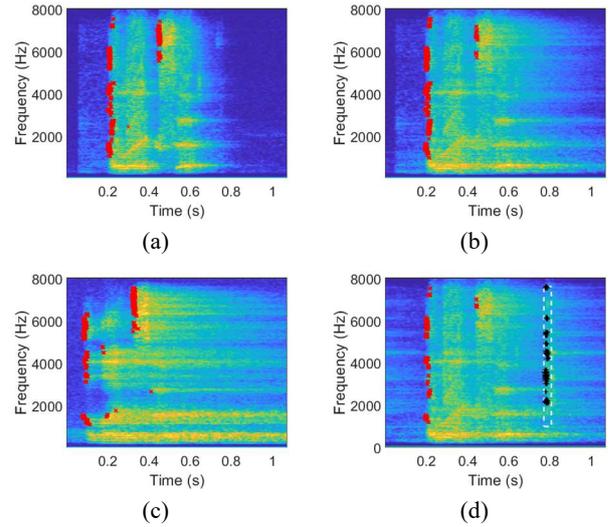

Figure 1: *The time-frequency bins passing the DPD test shown in red "×"s for a speech sample captured in (a) Room 1, (b) Room 2, (c) Room 3, (d) Room 2 in the presence of knocking interference with SIR = 5 dB. The transient noise is circled by white dotted lines. The time-frequency bins eliminated from the onset DPD test due to the transient noise determination are marked with black "+"s. The background is the distribution of the power envelope in the STFT domain.*

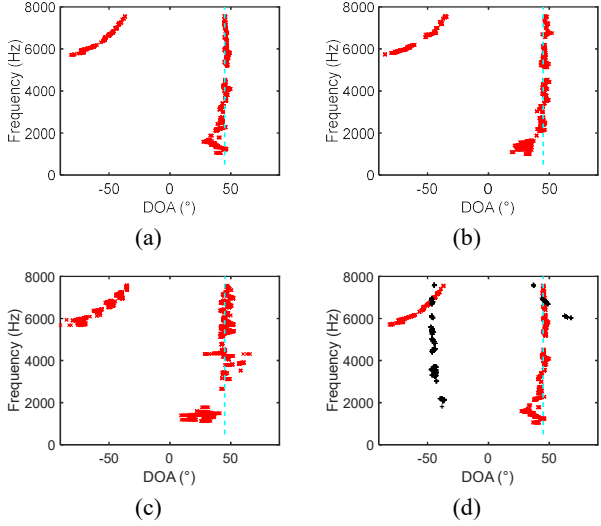

Figure 2: *The distribution of the estimated DOA for the selected bins in Fig. 1. The red "×"s represent the time-frequency bins that pass the DPD test, while the black "+"s represent the transient noise time-frequency bins that are eliminated from the DPD test. The cyan dashed line indicates the correct source direction.*

Figure 2 shows the distribution of the estimated DOA for the selected bins in Figure , in which the spatial aliasing can be clearly seen above the spatial Nyquist frequency of 4914 Hz for 0.035 m microphone interval. It can be seen from Figure 2 that the DOA estimated in the low-frequency band usually deviates considerably, and therefore the frequency bins which are under 1000 Hz or above 4914 Hz will also be neglected.

As shown in Figs. 1 and 2, in high reverberant environments, the set Π is highly likely to consist of the rising edge of the beginning of the speech since the rest of the speech tends to be flattened due to the influence of reverberation. The time-frequency bins in the set Π inevitably include those in the middle part of the speech in low reverberant environments. Fortunately, DOA estimation does not suffer significantly from reverberation in this condition and therefore accurate estimation result can still be expected.

While the transient interference is in existence, the determination and elimination of transient noises method work well, avoiding the influence of interference on target speeches DOA estimation.

Different people's voices have different characteristics in frequency. Some people have very low power in the high frequency part, especially in some male voices. In order to avoid the power of speech is much smaller than the power of interference in DOA estimation in high-frequency band, we must first locate in the middle-frequency band to obtain the general direction of the sound. And then, the precise DOA is estimated in high-frequency band.

Therefore, the set Π can be divided into two subsets as $\Pi_M$ and $\Pi_H$, and a two-stage DOA estimation scheme can be designed. Firstly, the set $\Pi_M$ can be exploited to develop a robust but possibly biased estimation of DOA. Then the refinement of the DOA can be obtained by making full use of the set $\Pi_H$.

## 2.2. DOA estimation in $\Pi_M$ with dereverberated bins

A proper dereverberation process is required in $\Pi_M$ since the middle frequency bins are still vulnerable to the influence of reverberation. The WPE algorithm is recognized as an effective method to eliminate late reverberation while preserving the direct and early reflections. However, the short frameshift of the onset DPD test conflicts the requirement of the normal WPE, in which usually one half or one fourth of the frame length is launched, and results in significant increase of the computational burden. A modified dereverberation scheme based on the WPE is proposed as follows.

The observed signal in set $\Pi_M$ can be decomposed into the desired signal $d_i(n,k)$ and the late reverberation $r_i(n,k)$ as

$$x_i(n,k) = d_i(n,k) + r_i(n,k), \qquad (10)$$

where $d_i(n,k)$ contains the direct speech signal and its early reflections. The late reverberation can be estimated from the observed signal by using the multi-channel linear prediction (MCLP) algorithm [26-28] and Estimate-Maximize (EM) algorithm [31]. The desired signal can be estimated as

$$d_i(n,k) = x_i(n,k) - \mathbf{w}_i^H(k)\mathbf{x}(n-D,k), \qquad (11)$$

where $\mathbf{w}_i(k)$ represents the prediction filter coefficients of channel $i$, $D$ the prediction delay, the superscript "H" denotes the conjugate transpose operation and $\mathbf{x}(n,k)=[x_1(n,k), x_2(n,k),…, x_I(n,k), x_1(n-p,k), x_2(n-p,k),…, x_I(n-p,k),…, x_1(n-(L-1)p,k), x_2(n-(L-1)p,k),…, x_I(n-(L-1)p,k)]^T$, with "T" denotes the matrix transposition operation, $L$ the prediction order in each subband, and $p$ the ratio between the frameshifts of the normal WPE and the DPD test.

WPE algorithm is used to estimate the desired signal. The cost function[29,30] based on the weighted least squares principle can be defined as

$$J(\mathbf{w}_i,k) = \sum_n \frac{|d_i(n,k)|^2}{\varepsilon_i^2(n,k)}, \qquad (12)$$

where $\varepsilon_i^2(n,k)$ is the variance of the desired signal.

The estimation procedure of the desired signal $d_i(n,k)$ is summarized as follows.

(1) Compute $\varepsilon_i^2(n,k)$

$$\varepsilon_i^2(n,k) = \max\left\{|x_i(n,k)|^2, \epsilon\right\}, \qquad (13)$$

where $\epsilon > 0$ is a lower bound of $\varepsilon_i^2(n,k)$.

(2) Repeat the following steps until the likelihood function reaches the maximum value.

  (a) Update $\mathbf{w}_i(k)$

$$\mathbf{w}_i(k) = \left(\sum_n \frac{\mathbf{x}(n-D,k)\mathbf{x}^H(n-D,k)}{\varepsilon_i^2(n,k)}\right)^\dagger \cdot \sum_n \frac{\mathbf{x}(n-D,k)x_i^*(n,k)}{\varepsilon_i^2(n,k)}, \qquad (14)$$

where "†" is the Moore-Penrose pseudo-inverse.

  (b) Update $d_i(n,k)$

$$d_i(n,k) = x_i(n,k) - \mathbf{w}_i^H(k)\mathbf{x}(n-D,k). \qquad (15)$$

(c) Update $\varepsilon_i^2(n,k)$

$$\varepsilon_i^2(n,k) = \max\left\{|d_i(n,k)|^2, J(\mathbf{w}_i,k)\right\}. \quad (16)$$

Note that the filter estimate as described in Eq. 11 is different from the normal WPE algorithm due to the different definition of $\mathbf{x}(n,k)$. The matrix included in the process is of a much lower order. The dereverberation operation is calculated only at the frequencies selected by the onset DPD test rather at all frequencies, further reducing the computational burden.

After calculating the dereverberated speech in the set $\Pi_M$, the DOA of each bin can be estimated using the SRP [8] as

$$\hat{\theta}(n,k) = \arg\max_{\theta}\left(\mathbf{g}(k,\theta)^H \mathbf{S}_d(n,k)\mathbf{g}(k,\theta)\right), \quad (17)$$

where $\mathbf{S}_d(n,k) = \mathbf{D}(n,k)\mathbf{D}(n,k)^H$, $\mathbf{D}(n,k) = [d_1(n,k), d_2(n,k), \ldots, d_I(n,k)]^T$, and $\mathbf{g}(k,\theta)$ denotes to the steering vector at the direction $\theta$, given by

$$\mathbf{g}(k,\theta) = \exp\left(-j\omega_k \frac{\mathbf{u}\sin\theta}{c}\right), \quad (18)$$

with $j$ the imaginary unit, $\omega_k$ the angular frequency, $\mathbf{u}$ the spatial vector of the microphones, and $c$ the speed of sound. The estimated DOA can be changed into the corresponding TDOA as

$$\hat{\tau}(n,k) = \frac{u_0 \sin\left(\hat{\theta}(n,k)\right)}{c}, \quad (19)$$

where $u_0$ is the distance between two adjacent microphones.

A quasi-histogram of the distribution of $\hat{\tau}(n,k)$ is utilized to estimate the DOA within the set $\Pi_M$. For every candidate TDOA $\tau$, a corresponding neighborhood is defined as

$$\Theta(\tau,k) \triangleq \left[\tau - \frac{\alpha}{k}\left(\frac{u_0}{c}+\tau\right), \tau + \frac{\alpha}{k}\left(\frac{u_0}{c}-\tau\right)\right], \quad (20)$$

where $\alpha$ determines the width of this neighborhood. Note that the neighborhood tends to have a wider range at low frequencies to mitigate the comparatively larger estimation error. Furthermore, the neighborhood inclines towards the broadside direction of the array, partially offset the impact of DOA estimation skewed to broadside [24]. The TDOA of the set $\Pi_M$ is determined by $\Theta(\tau,k)$ with the maximum distribution of $\hat{\tau}(n,k)$ as

$$\hat{\tau}_M = \arg\max_{\tau} \sum_{(n,k)\in\Pi_M} B\left(\hat{\tau}(n,k)\in\Theta(\tau,k)\right), \quad (21)$$

where $B(\cdot)$ is a standard Boolean function that returns 1 when the argument is "true" and returns 0 when the argument is "false".

### 2.3. Refinement of DOA estimation

The TDOA estimated in the set $\Pi_M$ is robust but possibly biased. To further refine the estimation, the selected bins in the set $\Pi_H$ is utilized. The estimated TDOA $\hat{\tau}_H(n,k)$ in each bin of $\Pi_H$ can be obtained through the same procedure as described in Sec. 2.2. Note that in circumstances with moderate reverberation time, the modified WPE algorithm can be skipped since the high frequency elements are usually not severely contaminated by reverberation.

To make full use of $\hat{\tau}_M$ estimated in Sec. II.B and reduce the impact of interference, the time-frequency bin with TDOA $\hat{\tau}_H(n,k)$ closer to $\hat{\tau}_M$ should be weighted more than the other bins, and therefore a Gaussian weighted function can be designed as

$$W(n,k) = \begin{cases} \frac{1}{2\pi\sigma}\exp\left[-\frac{(\hat{\tau}_H(n,k)-\hat{\tau}_M)^2}{2\sigma^2}\left(\frac{c}{2u_0}\right)^2\right], & |\hat{\tau}_H(n,k)-\hat{\tau}_M| \leq 3\sigma \\ 0, & |\hat{\tau}_H(n,k)-\hat{\tau}_M| > 3\sigma \end{cases}, (22)$$

where $\sigma$ is the standard deviation of the Gaussian function. The weighted SRP-PHAT algorithm is used to obtain the final DOA estimation as

$$\hat{\theta} = \arg\max_{\theta}\sum_{(n,k)\in\Pi_H} W(n,k)\frac{\mathbf{g}(k,\theta)^H \mathbf{S}_d(n,k)\mathbf{g}(k,\theta)}{\mathbf{D}(n,k)^H \mathbf{D}(n,k)}. \quad (23)$$

As a comparison, the MUSIC algorithm is also applied to the selected bins and finally the DOA is estimated by fusing the estimates from the different bins [18].

### 2.4. Flowchart of the whole algorithm

The entire DOA estimation algorithm, based on the direct path detection combined with dereverberation and frequency-divided refinement, is abbreviated as DPD-D-FR (PHAT) and summarized in Fig. 3. Firstly, the direct path time frequency bins are selected from the onset test with the elimination of the transient noise. The selected bins are divided into the middle frequency set $\Pi_M$ and the high frequency set $\Pi_H$. Then the modified WPE algorithm is launched to remove the late reverberation. The TDOA of each selected bin in $\Pi_M$ is estimated using Eqs. 17-19, and a robust but possibly biased TDOA is obtained using Eqs. 20 and 21. Finally, a more precise DOA can be obtained by further refined estimation using the TDOA gathered in $\Pi_H$, as depicted in Eqs. 22 and 23.

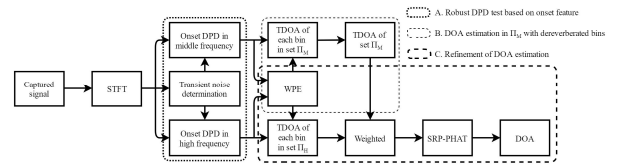

Figure 3: *The flow chart of DPD-D-FR (PHAT), with procedures described in Secs. 2.1 – 3 circled in dotted and dashed lines.*

## 3. Simulations

In this section, we present various simulations to illustrate the performance of the proposed algorithm. In all of the simulations, we employed a 6-channel ring array and adjacent interval of 0.035 m, which is easily installed on the top of cylindrical smart devices. The simulations were computed in rooms of the same size $7 \times 5 \times 3$ m$^3$ but different reverberation times: Room 1, T60 = 0.4 s; Room 2, T60 = 1 s. Room walls were assumed to have uniform reflection coefficients. The array center location was located at (3.5 m, 2.2 m, 1.5 m) and the array was placed horizontally in both rooms. The sound speed is c = 344 m/s. 15 piece of "ALEXA" wake-up words, 15 piece of "JINGLINGJINGLING" wake-up words and 20 piece of voices from the TIMIT database sampled by different speakers were respectively captured in 10 directions at intervals of 36° by the array with 2 m distance at the same height. When ambient interferences existed, 46 different disturbances are recorded at positions greater than 120° from the target speakers, also 2 meters from the center of the array, with SNR = 5 dB. The disturbances contain mechanical noise, music, fan noise, chain saw cutting noise, water boiling noise, market noise, ring tone, door slam, keyboard sound, rain sound, traffic noise, and other competing speaker's speech. The spherical harmonics coefficients of the plane waves density at the array position were synthesized using the image method [33] at a sampling frequency of 16 kHz. The ambient interferences continue throughout the process, and there are 0.2 s to 0.5 s of the target speaker blanks before and after the speeches in the waves we choose. All the parameters for the proposed algorithm are shown in Tab. 1. The portion below 1000 Hz is ignored in all algorithms.

Table 1: *Parameters for DPD-D-FR.*

| Parameters | Values |
|---|---|
| Window width | 512 |
| Frameshift | 8 |
| $\xi$ | $1 \times 10^{-3}$ |
| $N_t$ | 40 |
| $p$ | 32 |
| $K$ | The number of frames |
| $V_1, V_2$ | 3, 2 |
| $\Delta n$ | 72 |
| $\epsilon$ | $1 \times 10^{-4}$ |
| $\alpha$ | 8 |
| $\sigma$ | 1/15 |
| Range of middle frequency | [1000 Hz, 2000 Hz] |
| Range of higher frequency | [2000 Hz, 4914 Hz] |

The comparison in terms of the root-mean-square error (RMSE) of different algorithms in the absence of interference is shown in Fig. 2. In the presence of interference, we first counted the probabilities that the DOA estimation is closer to the interferences, denoted as $P_s$. Then the comparison in terms of RMSE of which DOA estimation is closer to the target speakers is also shown in Tab. 2, denoted as $R_s$.

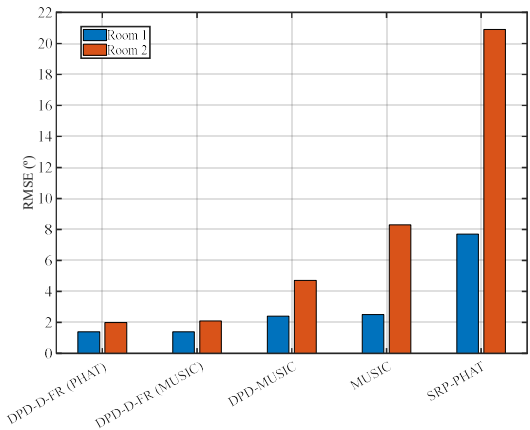

Figure 4: *RMSE of each algorithm in simulations.*

Table 2: $P_s$ *and* $R_s$ *of each algorithm in simulations.*

| Algorithm | Room 1 | | Room 2 | |
|---|---|---|---|---|
| | $P_s$ (%) | $R_s$ (°) | $P_s$ (%) | $R_s$ (°) |
| DPD-D-FR (PHAT) | 0 | 1.6 | 0 | 2.0 |
| DPD-D-FR (MUSIC) | 0 | 2.2 | 0.01 | 3.1 |
| DPD-MUSIC | 0 | 2.4 | 0.04 | 7.4 |
| MUSIC | 0 | 2.6 | 0.11 | 8.8 |
| SRP-PHAT | 3.89 | 9.0 | 17.9 | 22.0 |

It can be seen that the performance of the proposed algorithm is better than other algorithms. Influenced by the reverberation, the estimated DOA of SRP-PHAT severely deviates from the true incident angle. As a comparison, the algorithm we proposed is the most stable. Although there is little difference in the estimated DOA of ONSET-PHAT and ONSET-MUSIC, the amount of computation of ONSET-MUSIC required is much larger than ONSET-PHAT since MUSIC algorithm needs matrix eigenvector decomposition.

## 4. Experiments

In this section, we verified the performance of the proposed algorithm using the real captured audio in three rooms of Institute of Acoustics, Nanjing University, with different reverberation time. Room 1 is the auditory room with the size of $4.5 \times 7.4 \times 3$ m³ and T60 = 0.32 s, Room 2 is a small hall with the size of $3.6 \times 5.2 \times 3$ m³ and T60 = 1.20 s, and Room 3 is reverberation room with the size of $7.35 \times 5.9 \times 5.22$ m³ and T60 ≈ 5 s. A ULA with I = 4 microphones and adjacent interval of 0.035 m is used and its center is located at (1.5 m, 3 m, 1.5 m) in all rooms. The sound speed is c = 344 m/s. 15 piece of "ALEXA" wake-up words and 20 voices from the TIMIT database sampled by different speakers and 46 disturbances captured by the ULA with 2 m distance at a sampling frequency of 16 kHz. During the recording, the target speech source and the interference source are both 2 m away from the center of the array and placed at the same height. The interference sources are fixed at −45°, with SNR = 5 dB. The speaker sources are 30° and 60°.

The comparison in terms of the RMSE (°) of different algorithms in the absence of interference and in Room 2 with interference is shown in Tab. 3.

Table 3: *RMSE of each algorithm in experiments.*

| Room | 1 | | 2 | | 3 | | 2 (noise) | |
|---|---|---|---|---|---|---|---|---|
| Speaker (°) | 30 | 60 | 30 | 60 | 30 | 60 | 30 | 60 |
| DPD-D-FR (PHAT) | 0.6 | 1.0 | 0.7 | 0.7 | 0.7 | 1.2 | 1.1 | 1.1 |
| DPD-D-FR (MUSIC) | 0.7 | 1.6 | 0.8 | 0.8 | 0.8 | 3.0 | 1.3 | 1.6 |
| DPD-MUSIC | 0.9 | 1.8 | 1.1 | 1.0 | 1.0 | 3.6 | 1.3 | 1.7 |
| MUSIC | 1.0 | 1.9 | 1.1 | 1.0 | 1.0 | 3.6 | 1.6 | 3.0 |
| SRP-PHAT | 2.7 | 3.6 | 1.7 | 3.0 | 3.0 | 9.6 | 3.5 | 31 |

It can be also seen that the performance of the proposed algorithm is better than other algorithms. The performance of the DPD-D-FR (PHAT) is more stable under the influence of reverberation, with a maximum RMSE of 1.2° for the clean speech in all rooms. The interference also has a considerable negative effect on the estimation result. DPD-D-FR (PHAT) algorithm has similar RMSEs to other MUSIC-related algorithms, but the amount of computation is much smaller than those algorithms.

## 5. Conclusions

In this paper, a robust DOA estimation algorithm is proposed for the linear microphone array. The onset test is used to extract the time frequency bins of the direct speech signal, and an intuitive but effective scheme is used to ameliorate the influence of the transient noise. The two-stage DOA estimation procedure is designed based on the selected bins. The robust but possibly biased estimation is obtained in the middle frequency range using the dereverberated bins by the modified WPE algorithm, followed by further refinement in the high frequency in order to mitigate the deviation. The experiments in two typical reverberant environments with different interference noise level validates the efficacy of the proposed algorithm.

## 6. Acknowledgements

This work was supported by the National Natural Science Foundation of China (Grant No. 11874219).